# Automated Classification of First-Trimester Fetal Heart Views Using Ultrasound-Specific Self-Supervised Learning

1st Youssef Megahed
*Department of Systems and Computer Engineering*
*Carleton University*
Ottawa, Ontario, Canada
youssefmegahed@cmail.carleton.ca

2nd Aylin Erman
*Department of Clinical Science and Translational Medicine*
*University of Ottawa*
Ottawa, Ontario, Canada
aerma086@uottawa.ca

3rd Robin Ducharme
*Department of Acute Care Research*
*Ottawa Hospital Research Institute*
Ottawa, Ontario, Canada
roducharme@ohri.ca

4th Mark C. Walker
*Department of Acute Care Research*
*Ottawa Hospital Research Institute*
Ottawa, Ontario, Canada
mwalker@toh.ca

5th Steven Hawken
*Department of Methodological and Implementation Research*
*Ottawa Hospital Research Institute*
Ottawa, Ontario, Canada
shawken@ohri.ca

6th Adrian D. C. Chan
*Department of Systems and Computer Engineering*
*Carleton University*
Ottawa, Ontario, Canada
adrianchan@cunet.carleton.ca

*Abstract*—Congenital heart disease remains the most common congenital anomaly and a leading cause of neonatal morbidity and mortality. Although first-trimester fetal echocardiography offers an opportunity for earlier detection, automated analysis at this stage is challenging due to small cardiac structures, low signal-to-noise ratio, and substantial inter-operator variability. In this work, we evaluate a self-supervised ultrasound foundation model, USF-MAE, for first-trimester fetal heart view classification. USF-MAE is pretrained using masked autoencoding modelling on more than 370,000 unlabelled ultrasound images spanning over 40 anatomical regions and is subsequently fine-tuned for downstream classification. As a proof of concept, the pretrained Vision Transformer encoder was fine-tuned on an open-source dataset of 6,720 first-trimester fetal echocardiography images to classify five categories: aorta, atrioventricular flows, V sign, X sign, and Other. Model performance was benchmarked against supervised convolutional neural network baselines (ResNet-18 and ResNet-50) and a Vision Transformer (ViT-B/16) model pretrained on natural images (ImageNet-1k). All models were trained and evaluated using identical preprocessing, data splits, and optimization protocols. On an independent test set, USF-MAE achieved the highest performance across all evaluation metrics, with 90.57% accuracy, 91.15% precision, 90.57% recall, and 90.71% F1-score. This represents an improvement of +2.03% in accuracy and +1.98% in F1-score compared with the strongest baseline, ResNet-18. The proposed approach demonstrated robust performance without reliance on aggressive image preprocessing or region-of-interest cropping and showed improved discrimination of non-diagnostic frames. These results indicate that ultrasound-specific self-supervised pretraining enables more transferable and generalizable representations for early fetal cardiac imaging.

## I. Introduction

Congenital heart disease (CHD) is among the most common fetal abnormalities, yet early prenatal detection remains challenging [1]. The first trimester offers a crucial opportunity for screening, as identifying cardiac anomalies at 11-14 weeks can improve neonatal outcomes by enabling timely interventions [1]. However, first-trimester fetal echocardiography is technically demanding: the fetal heart is minute and rapidly moving, image quality is often limited, and the examination is highly operator-dependent [1-5]. As a result, sonographers' manual screening in early pregnancy can miss many cardiac defects (with low detection rates reported) [1]. There is a pressing need for automated decision support to enhance early fetal heart assessments.

Recent advances in deep learning (DL) have shown promise in prenatal ultrasound imaging. In particular, convolutional neural networks (CNNs) have been applied to classify standard fetal heart views from first-trimester ultrasound sweeps. For example, Stoean *et al.* [6] demonstrated that DL can effectively distinguish proper cardiac planes even at 12-14 weeks of gestational age, when the heart is still very small [1, 6]. Nonetheless, these pioneering works relied on careful image preprocessing to succeed. In particular, the usefulness of cropping and other image enhancement steps was explored to focus the CNN on the fetal heart region [1, 6]. While region-of-interest cropping can reduce background noise, it also risks discarding contextual cues and requires an accurate heart localization upfront. Moreover, first-trimester ultrasound videos contain a high proportion of "non-diagnostic" frames that do not show any standard view. Classifying such miscellaneous frames into an "Other" category is important for a realistic system (to avoid false positives), but it remains challenging because the Other class is inherently broad and imbalanced [1, 6]. Overall, conventional supervised models for this task have been constrained by limited annotated data, heavy preprocessing requirements, and potential generalization issues when applied to diverse scanning conditions.

To address these challenges, the field is turning to self-supervised learning (SSL) and foundation models for ultrasound imaging. Foundation models pretrained on large-scale medical data have achieved notable success in other domains, and ultrasound is beginning to show similar trends [5, 7, 8]. By learning from vast unlabeled ultrasound datasets, an SSL-based model can capture universal sonographic patterns (textures, anatomical shapes, artifacts) that traditional ImageNet-based models [9] might miss. Several groups have recently proposed ultrasound-specific foundation models. For example, Meyer *et al.*'s "UltraSam," trained on open ultrasound segmentation data [7], and Kang *et al.*'s URFM model for general ultrasound representation [8] highlight the growing interest in this area. Our team has also contributed to this paradigm by developing **USF-MAE (Ultrasound Self-**

**Supervised Foundation Model with Masked Autoencoding)**, the first large-scale MAE [10] foundation model pretrained exclusively on ultrasound data [5]. USF-MAE utilizes a Vision Transformer (ViT) encoder [11] and was pretrained via masked autoencoding on the OpenUS-46 corpus of over 370,000 ultrasound images spanning 46 anatomical categories [5]. This self-supervised pretraining allows the model to internalize diverse anatomical structures, scan planes, and imaging artifacts, thereby producing robust features for downstream tasks. In prior studies [12, 13], we showed that USF-MAE's learned representations transfer effectively to different prenatal ultrasound classification problems with minimal labels. For instance, our group demonstrated high diagnostic accuracy in detecting first-trimester cystic hygroma (a fluid-filled neck anomaly) using a DL model (DenseNet-169), achieving ~93% accuracy in distinguishing cystic hygroma from normal cases [12], and similarly applied CNNs to mid-trimester fetal renal scans for classifying congenital kidney anomalies [13]. These early works underscored the potential of AI in prenatal ultrasound while also revealing limitations of task-specific models, such as the need for consistent preprocessing and struggles with class imbalance. Building on these efforts, USF-MAE was shown to generalize across such tasks with improved efficiency [14, 15], indicating that a comprehensive ultrasound-pretrained model can overcome some shortcomings of bespoke CNN models.

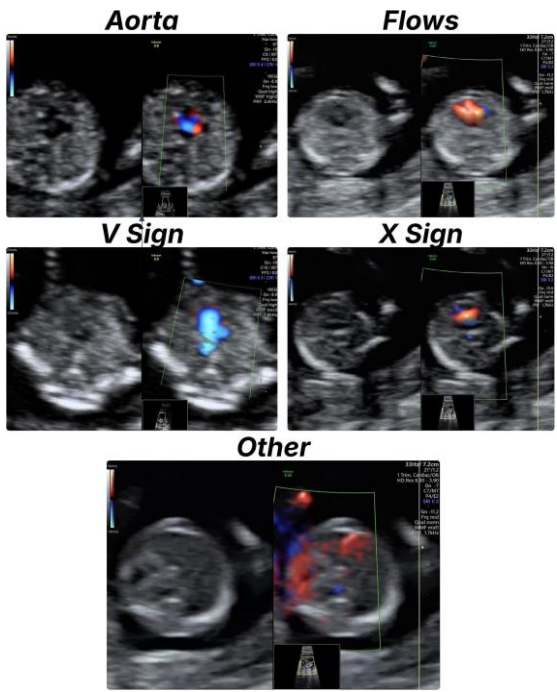

Figure 1: Ultrasound examples of fetal heart views showing the aorta "Aorta", the atrioventricular flows "Flows", the arches "V Sign", the crossing of the great vessels "X Sign", and the "Other" views. The "Other" class contains frames from the ultrasound videos that do not represent any of the other 4 specified key views.

In this paper, we present the first evaluation of an ultrasound foundation model for first-trimester fetal heart-view classification. We fine-tune the USF-MAE model on a labelled dataset of early fetal echocardiography images to automatically recognize the key standard cardiac views (the aorta "Aorta", the atrioventricular flows "Flows", the arches "V Sign", and the crossing of the great vessels "X Sign") versus the "Other" class (an example of each class is shown in Fig. 1). Unlike prior CNN-based approaches, our method leverages ultrasound-specific feature learning and does not require aggressive cropping or hand-crafted preprocessing of the input images. We explicitly investigate the impact of the challenging "Other" category and the model's ability to handle uncropped frames containing varying context. **To our knowledge, this is the first study to apply a self-supervised pretrained transformer model to early fetal heart scan analysis.** We rigorously compare the USF-MAE-based classifier against conventional DL baselines: a representative CNN (fine-tuned with ImageNet weights transfer learning) and a standard ViT without ultrasound pretraining. This comparative evaluation highlights the gains from domain-specific pretraining. We show that USF-MAE achieves higher classification accuracy and better generalization, particularly in correctly identifying the "Other" frames, than the baseline models. In summary, the contributions of this work include: (1) introducing an ultrasound foundation model approach for first-trimester fetal heart view recognition, (2) demonstrating improved performance and reduced need for manual preprocessing by using a USF-MAE transfer learning paradigm, and (3) providing a comprehensive comparison to CNN and ViT benchmarks to validate the effectiveness of ultrasound-tailored pretraining. By addressing the clinical motivation of early fetal cardiac screening with a novel technical solution, this study aims to advance the state of the art in automated fetal echocardiography and pave the way for more reliable early detection of cardiac anomalies.

II. METHODOLOGY

*A. Dataset and Problem Formulation*

This study focuses on the automated classification of first-trimester fetal heart ultrasound images into five clinically relevant categories (Fig. 1): aorta, atrioventricular flows (four-chamber view), V sign (three-vessel view), X sign (right ventricular outflow tract), and Other. The Other class comprises non-diagnostic frames that do not contain any of the standard cardiac views and represents the majority of frames encountered during routine fetal echocardiography sweeps [16].

We used the publicly available dataset introduced by Stoean *et al.* [16], which consists of 6,720 ultrasound images extracted from first-trimester fetal echocardiography video scans acquired between 12 and 14 weeks of gestation. Images were collected using multiple ultrasound machines and by sonographers with varying levels of expertise, reflecting realistic clinical variability. All images were manually annotated by expert clinicians and verified to ensure diagnostic correctness [16].

To prevent information leakage, all images originating from the same patient were assigned exclusively to a single data split. The dataset was divided into training, validation, and test subsets using a 50%, 25%, and 25% split, respectively, while preserving class distributions across splits.

*B. Image Preprocessing*

All images underwent a standardized preprocessing pipeline designed to remove patient-identifying information while preserving diagnostically relevant contextual information. First, fixed border cropping was applied to eliminate embedded patient metadata commonly present at the top of ultrasound images, as described in prior prenatal ultrasound studies [6, 16].

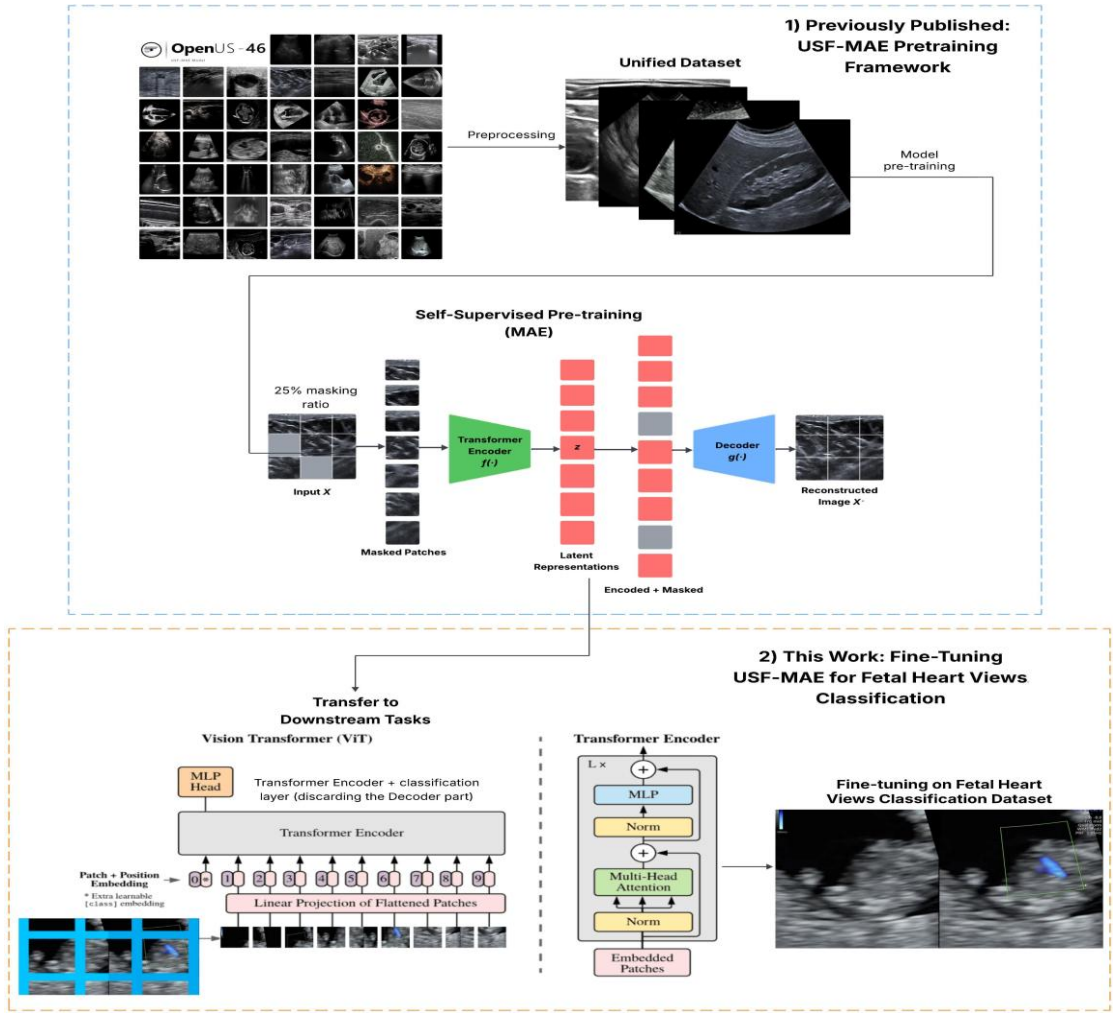

Figure 2: Overview of the proposed USF-MAE framework and its application to first-trimester fetal heart view classification. **Top:** Previously published USF-MAE self-supervised pretraining pipeline, where a unified ultrasound dataset (OpenUS-46) undergoes standardized preprocessing before MAE training. During pretraining, 25% of image patches are randomly masked and reconstructed using a lightweight decoder, while a ViT encoder learns general ultrasound representations [5]. **Bottom:** Downstream fine-tuning stage used in this work, where the pretrained MAE decoder is discarded and the transformer encoder is transferred to a fetal echocardiography classification task. A classification head is attached to the encoder, and the model is fine-tuned end-to-end to classify first-trimester fetal heart ultrasound images into five categories: aorta, atrioventricular flows, V sign, X sign, and Other.

Unlike prior approaches that rely on aggressive cropping of the fetal heart region [16], no convex hull extraction or region-of-interest cropping was applied in this work. This design choice was motivated by evidence that removing the surrounding anatomical context can negatively impact the classification of the Other class by eliminating informative background cues. Retaining full-frame images enables the model to learn both cardiac structures and contextual patterns associated with non-diagnostic frames.

All images were resized to $224 \times 224$ pixels and normalized using ImageNet [9] mean ([0.485, 0.456, 0.406]) and standard deviation ([0.229, 0.224, 0.225]) values to ensure compatibility across CNN and transformer-based architectures.

### C. USF-MAE Pretraining

The core of the proposed framework (Fig. 2) is the Ultrasound Self-Supervised Foundation Model with Masked Autoencoding (USF-MAE), a large-scale foundation model developed by our research group [5] and pretrained on unlabeled ultrasound data to learn general sonographic representations that transfer effectively across diverse downstream tasks. Unlike models pretrained on natural images, ultrasound imaging presents unique visual characteristics, including speckle noise, operator-dependent variation, and domain-specific texture patterns. These properties create a substantial domain gap when applying models pretrained on non-medical datasets such as ImageNet [9]. To address this, USF-MAE was specifically designed and pretrained to learn ultrasound-relevant features from a large corpus of unannotated sonographic images.

USF-MAE adopts the MAE SSL paradigm, a class of models that has demonstrated strong representation learning capability in vision tasks by reconstructing missing portions of the input from partially visible context [10]. The MAE framework consists of two major components: a high-capacity ViT encoder and a lightweight transformer decoder (Fig. 2). During pretraining, the encoder is tasked with modelling the semantics and structural regularities of ultrasound images, while the decoder reconstructs the missing content to enforce learning of informative latent representations.

Specifically, each ultrasound image is first resized to $224 \times 224$ pixels and partitioned into a grid of 196 non-overlapping patches of size $16 \times 16$. A random subset of these patches (25% in our model) is masked using a uniform sampling strategy, leaving the remaining 75% of patches visible to the encoder [5]. Positional embeddings are added to maintain the spatial relationships among the patches. The encoder processes the visible patches using multi-head self-attention mechanisms [17], capturing both local texture and long-range anatomical context.

The corresponding masked tokens are then concatenated with the encoded visible patch embeddings and passed to the lightweight decoder. The decoder's role is to predict the pixel values of the masked patches solely from the encoded visible context. Reconstruction quality is measured by the Mean Squared Error (MSE), shown in Eq. 1, loss computed only on the masked patch locations, which encourages the model to infer plausible ultrasound patterns based on learned anatomical and imaging priors [5]. This sparse reconstruction objective reduces the influence of irrelevant background and forces the encoder to focus on meaningful structural variations across the training corpus.

$$\mathcal{L}_{MAE} = \frac{1}{N} \sum_{i=1}^{N} \|\hat{x}_i - x_i\|_2^2 \quad (1)$$

USF-MAE was pretrained on OpenUS-46, a curated dataset of over 370,000 unlabelled ultrasound images gathered from 46 open-source datasets spanning more than 20 anatomical regions, including obstetric, abdominal, and musculoskeletal ultrasound [5]. This large and varied corpus allows the model to internalize a broad spectrum of sonographic features beyond any single clinical application. Pretraining on such diverse data supports cross-anatomical generalization, enabling the learned encoder to serve as a reusable backbone for many downstream classification tasks, including first-trimester fetal heart view recognition.

After pretraining, the decoder is discarded, and only the encoder remains for fine-tuning. By initializing downstream models with USF-MAE encoder weights rather than random or ImageNet weights, the framework capitalizes on ultrasound-specific representations that are better aligned with the statistical properties of the target ultrasound task [5]. Prior evaluations of USF-MAE on multiple ultrasound classification benchmarks, including breast lesion classification (BUS-BRA) [5], multiclass ovarian tumour classification (MMOTU-2D) [5], gastrointestinal tumour analysis (GIST514-DB) [5], fetal brain anomaly detection [2], cystic hygroma classification [14], and fetal renal anomaly detection [15], demonstrate superior performance over conventional CNN and transformer baselines when fine-tuned on limited labelled data.

### D. Fine-Tuning for Fetal Heart View Classification

For fetal heart view classification, the pretrained USF-MAE encoder was initialized with its self-supervised weights, and a fully connected multilayer perceptron classification head was attached to the encoder output. The entire network was fine-tuned end-to-end using labelled fetal echocardiography images.

Training was performed for 120 epochs using a batch size of 64. The AdamW [18] optimizer was employed due to its effectiveness in transformer-based models. A grid search was conducted over learning rates $\{3 \times 10^{-4}, 5 \times 10^{-4}, 1 \times 10^{-3}\}$ and weight decay values $\{0.01, 0.001, 0.0001\}$. The optimal configuration was found to be a learning rate of $5 \times 10^{-4}$ and a weight decay of 0.01, based on validation set performance. A cosine annealing learning rate schedule with linear warm-up was applied to stabilize early training and improve convergence. Cross-entropy loss was used as the optimization objective.

### E. Baseline Models

To provide a better context for USF-MAE's performance, three popular supervised baselines were deployed and trained under the same data splits and preprocessing settings. ResNet-18 and ResNet-50 are two popular CNNs of varying depth and capacity [19]. Both networks were initialized with weights pretrained on ImageNet [9] and fine-tuned on the fetal heart dataset. ViT-Base (ViT-B/16) [11] is a transformer architecture which was pretrained on ImageNet [9] and then fine-tuned for the five-class classification task. All baseline models were trained with the same optimizer, batch size, number of epochs, and data augmentation strategy.

### F. Data Augmentation and Regularization

To improve generalization, geometric data augmentation was applied only to training images. Augmentations included random rotations between 0° and 90°, horizontal and vertical flips with a probability of 0.5, and random resized cropping with scale factors between 0.5 and 2.0. No intensity-based augmentations were applied to avoid altering ultrasound-specific grayscale characteristics. Validation and test images were not augmented.

### G. Evaluation Metrics

Model performance was assessed using accuracy, precision, sensitivity (recall), and F1-score computed on both validation and independent test sets. Given the clinical importance of avoiding false-positive detections of diagnostic views, particular emphasis was placed on performance for the Other class. Additionally, one-vs-rest receiver operating characteristic (ROC) curves and precision–recall curves were generated for each class, and the area under the curve (AUC) was computed to assess discrimination performance across decision thresholds.

Table 1: Performance comparison (%) of the fetal heart views classification models on the test set.

| Model | Accuracy | Precision | Recall | F1-score |
|---|---|---|---|---|
| ResNet-18 | 88.54 | 89.41 | 88.54 | 88.73 |
| ResNet-50 | 86.95 | 88.08 | 86.95 | 87.26 |
| ViT-B/16 | 87.16 | 87.16 | 87.16 | 87.16 |
| **USF-MAE (ours)** | **90.57** | **91.15** | **90.57** | **90.71** |

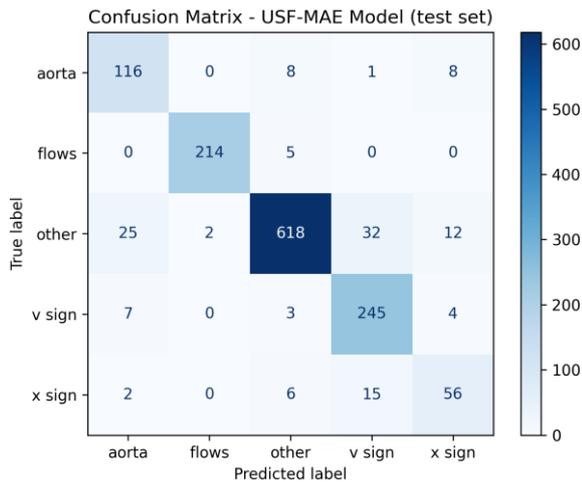

Figure 3: Confusion matrix of the USF-MAE model on the independent test set, showing classification performance across five first-trimester fetal heart view categories.

## III. RESULTS

Table 1 presents the quantitative performance of the proposed USF-MAE model and the baseline architectures on the independent test set for first-trimester fetal heart view classification. Across all evaluation metrics, USF-MAE achieved the highest performance, demonstrating consistent improvements over both convolutional and transformer-based baselines.

The USF-MAE model achieved a test accuracy of 90.57%, outperforming ResNet-18 (88.54%), ResNet-50 (86.95%), and ViT-B/16 (87.16%). Similar trends were observed for precision, recall, and F1-score. USF-MAE obtained a precision of 91.15%, a recall of 90.57%, and an F1-score of 90.71%, compared with F1-scores of 88.73% for ResNet-18, 87.26% for ResNet-50, and 87.16% for ViT-B/16.

Notably, despite ViT-B/16 sharing a similar transformer-based architecture, its performance remained inferior to USF-MAE across all metrics, highlighting the benefit of ultrasound-specific self-supervised pretraining over transfer learning from natural images alone. The deeper ResNet-50 model also did not outperform the shallower ResNet-18, suggesting that increased model capacity alone does not guarantee improved performance in data-limited first-trimester ultrasound tasks.

Overall, the results demonstrate that USF-MAE provides superior generalization on unseen test data when compared under identical training, preprocessing, and evaluation conditions. These findings support the effectiveness of large-scale self-supervised ultrasound pretraining for improving downstream fetal echocardiography classification performance.

## IV. DISCUSSION

### A. Key Findings

The results of this study demonstrate that self-supervised, ultrasound-specific foundation models provide a clear advantage over conventional supervised DL approaches for first-trimester fetal heart view classification. The proposed USF-MAE model consistently outperformed all baseline models, including ResNet-18, ResNet-50, and ViT-B/16, across accuracy, precision, recall, and F1-score on the independent test set. Notably, this performance gain was achieved under identical preprocessing, training, and evaluation conditions, indicating that the observed improvements arise from the quality of the learned representations rather than from differences in optimization or data handling.

These findings are consistent with prior work from our group [2, 14, 15], where USF-MAE demonstrated superior performance compared to supervised CNN baselines, including DenseNet-based models [13, 20], across multiple prenatal ultrasound applications such as cystic hygroma detection [14], fetal renal anomaly classification [15], and fetal neurosonography [2]. The reproducibility of these gains across distinct anatomical targets and datasets supports the central hypothesis that large-scale self-supervised pretraining on heterogeneous ultrasound data enables more transferable and robust feature learning. The USF-MAE framework and pretrained weights are publicly available at our GitHub repository "https://github.com/Yusufii9/USF-MAE" to promote transparency and reproducibility.

An important observation from this study is that improved performance was achieved without relying on aggressive image cropping or region-of-interest extraction. Prior work on the same fetal heart dataset [6, 19] showed that preprocessing steps aimed at isolating the cardiac region can inadvertently degrade performance, particularly for the clinically important Other class, by removing contextual information necessary to distinguish diagnostic from non-diagnostic frames. In contrast, USF-MAE effectively leverages full-frame images, suggesting that the pretrained encoder learns to attend to relevant anatomical structures while still exploiting global context. This reduced dependence on heuristic preprocessing improves robustness and simplifies deployment in real-world clinical settings.

Analysis of the confusion matrix (Fig. 3) further illustrates the strengths and remaining challenges of the proposed approach. High true positive rates were observed for all five classes, with particularly strong performance for the atrioventricular flows and V sign views. Misclassifications were primarily concentrated between anatomically related views and the Other class, reflecting inherent ambiguity in certain first-trimester frames rather than systematic model failure. Importantly, the majority of Other frames were correctly classified, addressing a key limitation of earlier CNN-based approaches that struggled with false positive detection of standard views.

The ROC and precision-recall curves (Fig. 4) provide additional insight into class-wise discrimination performance. High area under the ROC curve values across all classes indicate strong separability over a wide range of decision thresholds. Precision-recall analysis highlights robust performance under class imbalance, particularly for the Other class, which constitutes a large proportion of the dataset. The comparatively lower performance observed for the X sign class is consistent with its visual subtlety and lower prevalence in early gestation ultrasound, yet the model still demonstrates meaningful predictive capability for this challenging category.

### B. Clinical Implications

The results of this study are of clinical interest in advancing a more robust, earlier approach to fetal cardiac screening in the first trimester. Reliable identification of normal fetal heart views is a prerequisite for an automated

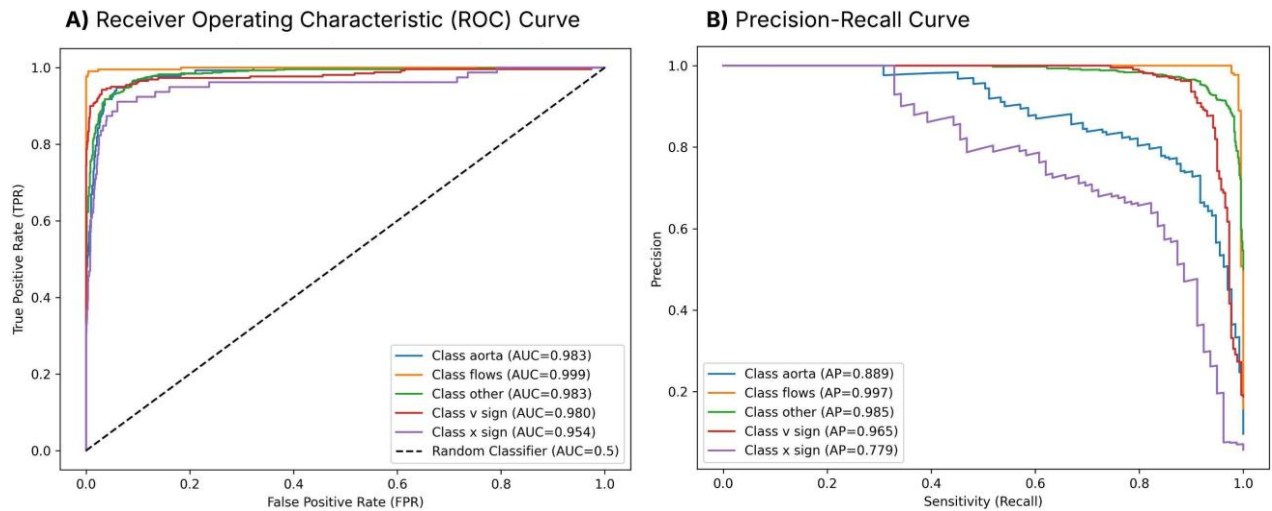

Figure 4: ROC and precision-recall curves of the USF-MAE model on the independent test set, reported in a one-vs-rest setting for each fetal heart view class.

CHD screening system, and improvements in the initial step of the process may enable earlier referral, follow-up imaging, and perinatal care planning. The proposed approach may also reduce false positives and improve the identification of non-diagnostic frames, potentially addressing the issue of false reassurance arising from incomplete cardiac scans mistakenly interpreted as normal.

The reduced reliance on manual preprocessing steps and region-specific heuristics may also improve the suitability of this model for real-time sonographer-assistance tools. A model which generalises to various acquisition conditions and can operate directly on full-frame ultrasound images is well placed for implementation as a clinical routine, where it could assist in standardising acquisition and providing feedback to the operator in real time as the scan is being performed, potentially improving completeness and consistency of examination by alerting the sonographer when required views have been captured, or when further sweeps are required.

*C. Limitations*

While these results are encouraging, we note that our work has several limitations. First, our current study is based on individual image frames from ultrasound videos. While frame-based classification is an important first step, fetal echocardiography is an inherently temporal modality, and fusing spatiotemporal information from ultrasound video sequences may improve view recognition and anomaly detection performance.

Second, our dataset comprises images from multiple ultrasound machines and operators, but vendor diversity was not explicitly controlled or analyzed. Performance may vary due to differences in image appearance across manufacturers, and future work should assess generalization performance on vendor-specific data subsets and on multi-institutional cohorts.

Future work will also aim to address these limitations, particularly as ultrasound foundation models scale and are leveraged for more complex clinical tasks.

## V. CONCLUSION

In this work, we presented a self-supervised ultrasound foundation model, USF-MAE, for first-trimester fetal heart view classification and demonstrated its effectiveness compared with conventional convolutional and transformer-based baselines. By leveraging large-scale self-supervised pretraining on heterogeneous ultrasound data, USF-MAE achieved consistently superior performance across accuracy, precision, recall, and F1-score on an independent test set, highlighting the benefit of ultrasound-specific representation learning over transfer learning from natural images (e.g., ImageNet).

The results show that robust fetal heart view classification can be achieved without reliance on aggressive image preprocessing or region-of-interest cropping, thereby preserving contextual information that is critical for distinguishing diagnostic from non-diagnostic frames. This characteristic enhances model robustness and supports practical deployment in real-world clinical environments. The improved performance observed across multiple fetal ultrasound applications in prior work further underscores the generalizability of the USF-MAE framework.

From a clinical standpoint, the proposed approach has the potential to support earlier and more reliable fetal cardiac screening by assisting sonographers during first-trimester examinations and reducing false positive detection of standard cardiac views. Such capabilities could contribute to improved early identification of congenital heart disease and more consistent screening quality.

Future work will focus on extending the framework to spatiotemporal video-based modelling, evaluating cross-vendor and multi-institutional generalizability, and exploring integration into real-time ultrasound acquisition systems. To promote transparency, reproducibility, and further research in ultrasound foundation models, the USF-MAE framework and pretrained weights are publicly available at our GitHub repository: **https://github.com/Yusufii9/USF-MAE**.